\documentclass[12pt,twoside]{article}
\usepackage{graphics}
\usepackage{fleqn,espcrc1}
\usepackage{epsfig}

\def\lsim{\mathrel{\rlap{\lower4pt\hbox{\hskip1pt$\sim$}}
    \raise1pt\hbox{$<$}}}         
\def\gsim{\mathrel{\rlap{\lower4pt\hbox{\hskip1pt$\sim$}}
    \raise1pt\hbox{$>$}}}         

\def\be{\begin{equation}}
\def\ee{\end{equation}}
\def\bq{\begin{eqnarray}}
\def\eq{\end{eqnarray}}

\begin{document}


\begin{titlepage}

\title{Gamma Ray Bursts via  emission  
of axion-like particles}

\author{
Zurab Berezhiani\address{
Dipartimento di Fisica, Universit\`a di L'Aquila, 
I-67010 Coppito (AQ) \\
and INFN, Laboratori Nazionali del Gran Sasso,  
I-67010 Assergi (AQ), Italy }
\address{Institute of Physics, Georgian Academy of Sciences, 
380077 Tbilisi, Georgia} 
and
Alessandro Drago\address{
Dipartimento di Fisica, Universit{\`a} di Ferrara \\
and INFN, Sezione di Ferrara, I-44100 Ferrara, Italy
}}

\maketitle
\bigskip
\bigskip
\bigskip
\bigskip

\begin{abstract}
 
The Pseudo-Goldstone Boson (PGB) emission could provide 
a very efficient mechanism for explaining the cosmic 
Gamma Ray Bursts (GRBs). The PGBs could be produced  
during the merging of two compact objects like 
two neutron stars or neutron star - black hole, 
and then decay into electron-positron pairs and photons  
at distances of hundreds or thousands km, where the
baryon density is low. In this way, a huge energy 
(up to more than $10^{54}$ erg)  
can be deposited into the outer space  
in the form of ultrarelativistic $e^+e^-$ plasma, the so called  
fireball, which originates the observed gamma-ray bursts. 
The needed ranges for the PGB parameters are: mass of order MeV, 
coupling to nucleons $g_{aN}\sim {\rm few}\times 10^{-6}$  
and to electrons $g_{ae}\sim {\rm few}\times 10^{-9}$. 
Interestingly enough, the range for coupling constants correspond 
to that of the invisible axion with the Peccei-Quinn symmetry 
breaking scale $f\sim {\rm few}\times 10^5$ GeV, but the mass of the
PGB is many orders of magnitude larger than what such a scale would 
demand to an axion.  
Neither present experimental data nor astrophysical and cosmological 
arguments can exclude such an ultramassive axion,  
however the relevant parameters' window is within 
the reach of future experiments. 
Another exciting point is that our mechanism could explain 
the association of some GRBs with supernovae type Ib/c, 
as far as their progenitor stars have a radius $\sim 10^4$ km. 
And finally, it also could help the supernova type II explosion:  
PGB emitted from the core of the collapsing star 
and decaying in the outer shells would deposit a 
kinetic energy of the order of $10^{51}$ erg.  
In this way, emission of such an axion-like particle could 
provide an unique theoretical base for understanding 
the gamma ray bursts and supernova explosions. 

\end{abstract}

\begin{flushright}
DFAQ-TH/99-05 \\ 
INFN-FE/12-99 
\end{flushright}

\vfill
\newpage
\end{titlepage}

\section{Introduction}

The phenomenon of the so-called Gamma-Ray Bursts (GRBs) puzzles the 
theorists from many points of view (for a review see \cite{piran}). 
The most striking feature of GRBs is that an enormous energy 
is released in few seconds in terms of gamma-rays
having typical energies of few hundred keV.  
The energy emitted in GRBs is  
up to $10^{53-54}$ erg, 
making it difficult to devise a mechanism
able to transform efficiently enough the gravitational energy into   
such a powerful photon emission.  
In addition, the bursts show a variety
of complex time-structures of the light curves. 
The time-structure of the prompt emission and the recent 
discovery of the afterglow fit the expectations 
of the so-called fireball model \cite{meszaros}, 
in which the electromagnetic radiation is originated from
the electron-positron plasma that expands at 
ultrarelativistic velocities undergoing internal
and external shocks. 
The Lorentz factor of the plasma has to be very large, 
$\Gamma\sim 100$, in order to solve the so-called 
compactness problem \cite{compact}. 
So large values of $\Gamma$ are difficult to achieve 
because they require a very efficient
mechanism to accelerate the plasma. In addition, 
in order to avoid the contamination of the $e^+ e^-$ plasma 
by more massive matter, the plasma
has to be produced in a region of low baryonic density.
  
Several mechanisms have been proposed to power the
$e^+ e^-$ plasma.  In particular, it could be obtained
via annihilation  $\nu\bar{\nu}\to e^+ e^-$
of the neutrinos emitted by a heavy compact
collapsing object -- 
collapsar \cite{Venya,paczynski,woosley99}. 
Alternatively, neutrinos
can be produced at the merger of two compact objects,
e.g. of two neutron stars \cite{eichler} or a neutron star
and a black hole \cite{bp}.
The latter possibilities were recently analyzed in
details in refs. \cite{ruffert,popham,janka99}.   
\footnote{Recently some exotic mechanisms
have been suggested related to the gravitational collapse of
mirror stars \cite{blinnikov}.}

The main challenge to the existing models is that the
energy released in photons is astonishingly large. 
For example,   
the Gamma-Ray Burst GRB 990123 shows an energy release 
${\cal E} \simeq 3.4\times 10^{54}$ erg
$ = 1.9 M_\odot$, if
isotropic emission is assumed.
Taking into account a possible beaming, this energy
can be reduced down to
${\cal E} \simeq 6\times 10^{52}$ erg \cite{kulkarni}.
There are also few other events (e.g. GRB 971214 and GRB 980703)
with typical energies ${\cal E}\sim 10^{53}$ erg, which show
no evidence for collimation \cite{harrison}.
The models invoking the $\nu\bar\nu \to e^+ e^-$ reaction
as a source for the GRB have strong difficulties in
reaching so large photon luminosities.
Although during the collapse of compact
objects a relevant amount of energy is normally emitted
in terms of neutrinos, the low efficiency of the 
$\nu\bar\nu \to e^+ e^-$ reaction 
strongly reduces the 
energy deposited in the fireball.

In the present paper we propose that
light pseudoscalar particles -- Pseudo-Goldstone Bosons (PGBs) --
can be extremely efficient messengers for transferring the
gravitational energy released in the merger of compact stars,
into ultrarelativistic $e^+e^-$ plasma.
We show that for certain parameter ranges
the PGBs can be effectively produced inside the dense
core of the collapsing system and then decay into $\gamma\gamma$
or $e^+e^-$ outside the system, in baryon free zones, 
thus giving rise to 
the ultrarelativistic $e^+ e^-$ plasma.

The advantage of assuming that the fireball is produced
via the PGB decay instead of 
$\nu\bar\nu\to e^+e^-$ annihilation is obvious. 
The latter process has a low efficiency -- neutrinos
deposit to plasma only few percent of the emitted energy 
and take the rest away.  
In addition, the process $\nu\bar\nu\to e^+e^-$
can be effective only at small distances, less than 100 km, 
which are still contaminated by baryon load, and it fails 
to provide a sufficiently large Lorentz factor to the plasma, 
at most $\Gamma\sim 5$ \cite{ruffert,popham}. 
Instead, the PGB mechanism is 100 percent efficient:
decay can take place at distances about 1000 km and, 
in addition, all energy emitted in terms of PGBs
is deposited to the $e^+e^-$ plasma with
the Lorentz factor $\Gamma \sim E/m_e$,
where $E$ is a typical energy of emitted PGBs (tens of MeV)     
and $m_e$ is the electron mass.
This analysis is supported by the
very recent result of ref. \cite{aloy}, which shows 
that if the energy would be transfered to the plasma
at distances of at least few hundreds km, a successful burst could
be obtained with a Lorentz factor $\sim 40$ or so. 

The most familiar example of a PGB, the axion \cite{PQ,kim}, 
has already been considered in connection with GRBs \cite{loeb}. 
However, this was an invisible axion with the 
mass $m_a \sim 10^{-5}$ eV.  
Such an axion has a lifetime much larger than the age of the universe 
and thus its decay is impossible. Instead, it was assumed that 
the axions could convert into photons in strong magnetic fields 
near the collapsing system. It is questionable, however, 
how large the magnetic field has to be in order to achieve 
a large efficiency for the conversion. 

In our analysis we phenomenologically consider a more general
possibility, assuming that the PGB mass $m_a$ and its couplings to  
nucleons, electrons and photons 
are not constrained by the Peccei-Quinn relations. 
The paper is organized as follows. In section 2 we shortly review the
results of refs. \cite{ruffert,popham,janka99} 
which describe the general scenario for mergers.
In section 3 we consider
the PGB emission as a source for the GRB
and discuss the ranges for its mass and couplings  
which are compatible with
the existing astrophysical constraints.
In section 4 possible models for such axion-like particles are discussed.
In section 5 we analyze possible signatures of the proposed mechanism.
In particular, we show that 
the PGBs emission could generate a GRB associated with the 
supernovae type Ib,c, and also help the supernova type II  
explosion. Finally, in section 6 we summarize our findings.  

\section{NS-NS and NS-BH mergers}

The scenarios depicted in refs. \cite{ruffert,popham,janka99} for 
the NS-NS and NS-BH mergers are actually very similar. 
In both cases an axially symmetric 
structure develops. In the center the system collapses into a 
black hole of a few solar masses,  while 
a fraction of matter ($M\sim 0.1 M_\odot$ for NS-NS and  
$M\sim 0.5 M_\odot$ for NS-BH) obtains enough angular momentum 
to resist immediate collapse into the black hole and remains
in an accretion torus around the black-hole. 
Then this mass accretes from the torus into the BH going through a disk-like
structure having a thickness which decreases approaching the BH.
The central part of the disk is the most relevant for the radiation
of energy, being the reservoir of largest density and temperature. 
The accretion rate from the torus $dM/dt$ is of the order of 
1 $M_\odot$/s for NS-NS and 5 $M_\odot$/s for NS-BH,  
so the duration of this phase 
is essentially similar in both cases, $t\sim 0.1$ s.

According to refs. \cite{ruffert,popham,janka99},  
the energy is radiated from the disk via the neutrino emission. 
The deposition of energy in a region with low baryonic mass proceeds 
through $\nu\bar\nu$ annihilation into $e^+ e^-$.
The observed GRB is produced by the $e^+e^-$ plasma which expands
at ultra-relativistic velocity. 
The energy deposited into the $e^+e^-$ plasma ranges from 
$E_{\nu\bar\nu}\sim 10^{48-49}$ erg in the case of NS-NS merger up to
$E_{\nu\bar\nu}\sim 10^{51.6}$ erg for the NS-NH merger. The Lorentz
factor are always rather small, $\Gamma\sim 5$. 

An important point in the simulations of refs. \cite{ruffert,popham}
concerns the density and temperature profiles of the system. 
In ref. \cite{ruffert}, where the
neutrino trapping has been taken into account, 
the maximum density is $\sim 10^{11}$ g/cm$^3$, with small regions reaching
$10^{12}$ g/cm$^3$.  
On the contrary, in ref. \cite{popham} where
neutrinos are assumed to skip freely, 
densities are generally smaller by about an
order of magnitude.
When neutrino trapping is taken into account, larger densities,
similar to the ones in ref. \cite{ruffert}, are obtained.
In ref. \cite{popham} it is indeed shown that the density of the disk
depends on the amount of energy lost by matter in this region. It is clear
that only a totally self-consistent calculation can indicate if neutrinos
are trapped, reducing the luminosity of neutrino emission and therefore
allowing for a larger density in the center of the disk. The connection
between cooling and density is due to the contribution to the pressure
coming from radiation and relativistic electrons and positrons. If the
cooling is switched-off, only gas pressure is present and the system
stabilizes at larger densities.

It will be important in our discussion to keep in mind that a more efficient
way of extracting energy from the system, e.g. via PGBs emission, will in turn
modify the density in the center of the disk. Since in our work we do not
pretend to solve the dynamics of the system, but we only indicate a new 
possibility for powering a strong GRB, we will consider for the central
density of the disk values ranging from $\sim 10^{11}$ g/cm$^3$ down
to $10^9$ g/cm$^3$.

The estimated temperature profiles of the disk
are rather similar in refs. \cite{ruffert}
and \cite{popham}, reaching about 10 MeV in the central part
and remaining above 1 MeV up to distances of the order of 100 km from
the center of the system.

\section{GRB via PGB emission}

Let us consider the following Lagrangian describing 
the PGB couplings to the fermions $\psi_i$ ($i=e,p,n,...$) and 
electromagnetic field-strength tensor $F_{\mu\nu}$: 
\be
{\mathcal L} = -\frac12 m^2_a a^2  
- ig_{ai}a \bar\psi_i\gamma_5\psi_i 
- {g_{a\gamma}\over 4m_N}a F_{\mu\nu}\tilde F^{\mu\nu} + .... 
\label{lagrangian} 
\ee
where in the last term the nucleon mass $m_N$ 
is taken as a regulator scale. 
In cases of the familiar models of the PGB like the axion,  
its mass and couplings 
are all determined, but for some coefficients, by only
one free parameter $f$, the scale of the global $U(1)_{PQ}$
symmetry.  
Instead, here we take a more phenomenological approach, 
considering $m_a$, $g_{a\gamma}$, $g_{aN}$ and $g_{ae}$ 
as independent quantities.

Let us start by discussing which 
constraints should  be imposed 
immediately on the PGB mass and its couplings 
in order to obtain a strong GRB without violating the already 
existing experimental and astrophysical limits. 
In particular, we consider the mass and the coupling to the
nucleon within the range 
\be
0.3~{\rm MeV} < m_a < {\rm few ~ MeV} 
\label{m-limit}
\ee
\be
10^{-6} < g_{aN} < 10^{-4}  
\label{g-limit}
\ee
The upper mass limit comes from the following. In order to 
have efficient PGB production 
in the dense systems under consideration, 
its mass should not over-exceed the characteristic temperature 
of the latter, typically of few MeV. 
The lower mass limit in 
(\ref{m-limit}) enables to avoid the astrophysical constraints 
from the stellar evolution. The PGB 
production rate in the stellar cores 
with typical temperatures $T$ up to 10 keV, 
is proportional to $g_{aN}^2 \exp(-m_a/T)$. 
Thus, in order to avoid too fast stellar cooling, 
the exponential factor needs to be small  
if the constant $g_{aN}$ falls in the range indicated 
in (\ref{g-limit}).
The lower limit on $g_{aN}$ comes from supernovae. It implies 
that the PGBs are trapped in the collapsing core 
and thus the SN 1987A neutrino signal is not affected 
\cite{raffelt}.\footnote{The region 
$g_{aN}\le 10^{-11}$ is also acceptable for the constraints 
coming from neutrino emission in supernovae \cite{raffelt}, but
is not of interest for our analysis. }
Finally, the upper limit in (\ref{g-limit}) stands for a  
conservative interpretation of the set of 
constraints coming from the search of the decay $K^+\to \pi^+ a$, 
from the beam dump and other terrestrial experiments \cite{rpp}.  
The experimental limits on the couplings $g_{ae}$ and $g_{a\gamma}$  
will be discussed later. 

Let us discuss now in more details the range of the PGB 
parameters needed for successfully producing the GRBs.

\subsubsection*{ Mean lifetime} 

The PGB can decay into photons and, if $m_a > 2m_e$, 
also into $e^+e^-$. The corresponding decay width is 
$\Gamma_{\rm tot}= \Gamma(a\to \gamma\gamma) + \Gamma(a\to e^+e^-)$, 
where
\be
\Gamma(a\to \gamma\gamma)
= {g_{a\gamma}^2\over 64\pi m_N^2} m_a^3, 
\ee 
\be
\Gamma(a\to e^+e^-)
= {g_{ae}^2\over 8\pi} m_a \left(1-\frac{4m_e^2}{m_a^2}\right)^{1/2} . 
\ee

As we anticipated in the introduction, we discuss the emission of PGBs
by the central part of the toroidal system with typical size $\sim 100$ km, 
obtained during the merger 
of two compact objects. 
We assume that these PGBs take away a large fraction of the energy
of the system and 
we want them to decay 
into photons or $e^+e^-$ preferentially 
outside the disk, in the region of low baryon density,  
thus giving rise to the hot relativistic plasma. 
Therefore a reasonable choice for the decay length of the PGBs,  
$D_a = c\tau\gamma$, is in the range
\be
100~{\rm km}< D_a < 10000~{\rm km}.
\label{distance} 
\ee 
Here $\tau=\Gamma_{\rm tot}^{-1}$ is the lifetime of the PGB 
at rest and $\gamma=E/m_a$ is its Lorentz factor. 
The average energy of the emitted PGBs can be taken 
as $E\simeq 3T$, where $T$ is a temperature of the system. 
Since $T$ is of the order of several MeV, the Lorentz factor 
$\gamma$ ranges from 2-3 up to 20 or more, 
depending on the mass of the PGB. 
The values of $g_{a\gamma}$ and $g_{ae}$ corresponding to a 
given decay length $D_a$,  
as functions of $m_a$, are shown in Fig. 1.  

The upper limit in (\ref{distance}) is not rigidly determined and 
is inferred from the following. 
In the literature it is argued that there is a correlation 
between some GRBs and supernovae type Ib/c 
\cite{kulkarni98,reichart}. On the other hand, 
the observed frequency of supernova explosions is 
orders of magnitude larger than the observed 
frequency of GRBs. Therefore not every SN explosion 
is accompanied by a detected GRBs. If a supernova type Ib/c emits
a GRB, the latter must be a rather weak one, so that it is detected only
in few cases. Therefore, $D_a$  should be less 
than the typical radius of the type Ib/c supernova progenitors 
($R\sim 10^4$ km), so that the strength of the associated 
GRB is suppressed by a factor $\exp(-R/D_a)$.  
Obviously, this suppression is enormous in the case of SN type II 
which progenitor has a typical radius larger than $10^7$ km. 
In particular this makes clear the absence of a detectable GRB associated with 
SN 1987A.

\begin{figure}

\parbox{6cm}{
\scalebox{0.6}{
\includegraphics*[80,445][440,700]{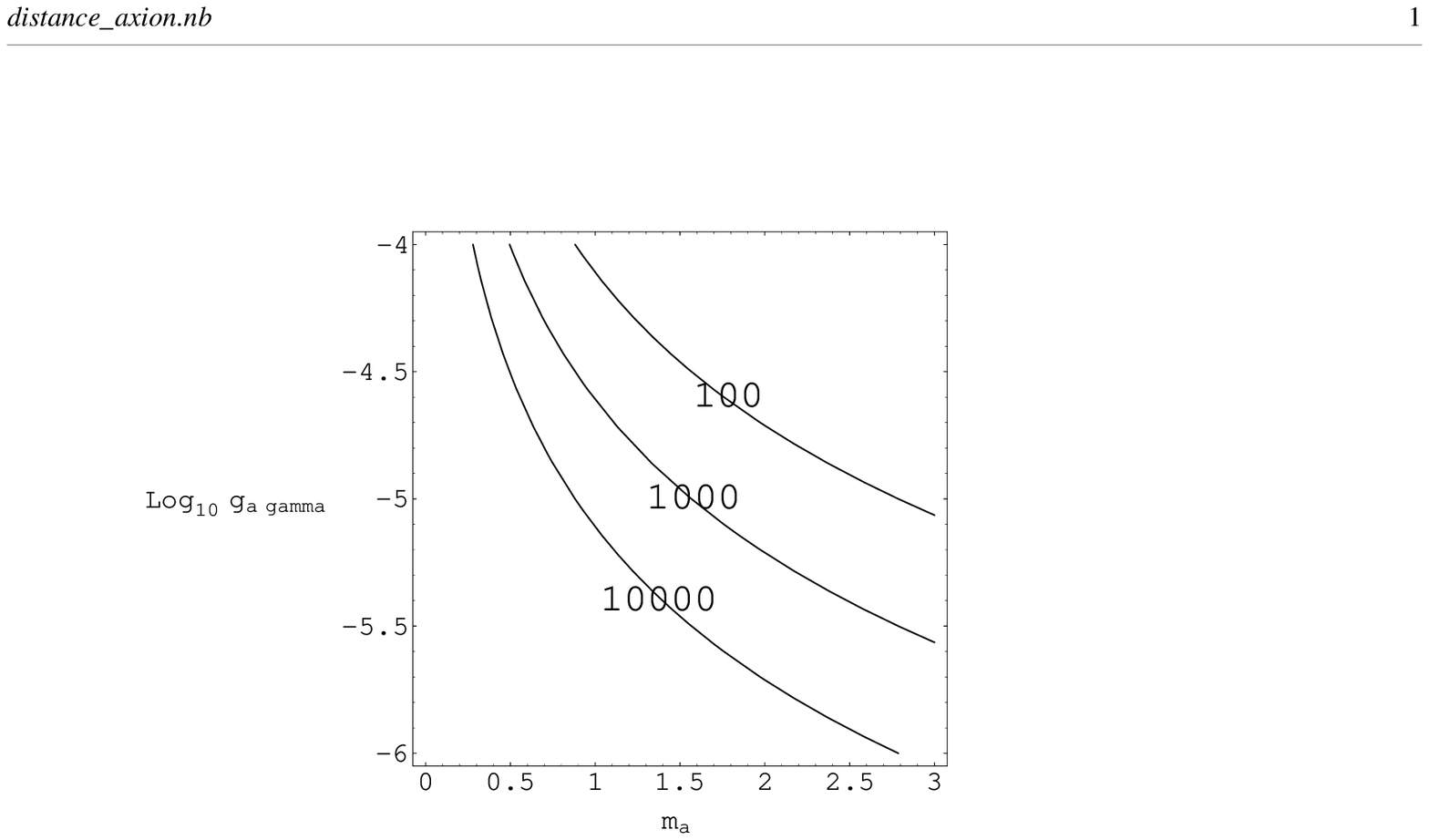}}
}
\nolinebreak
\parbox{6cm}{
\scalebox{0.6}{
\includegraphics*[40,445][410,700]{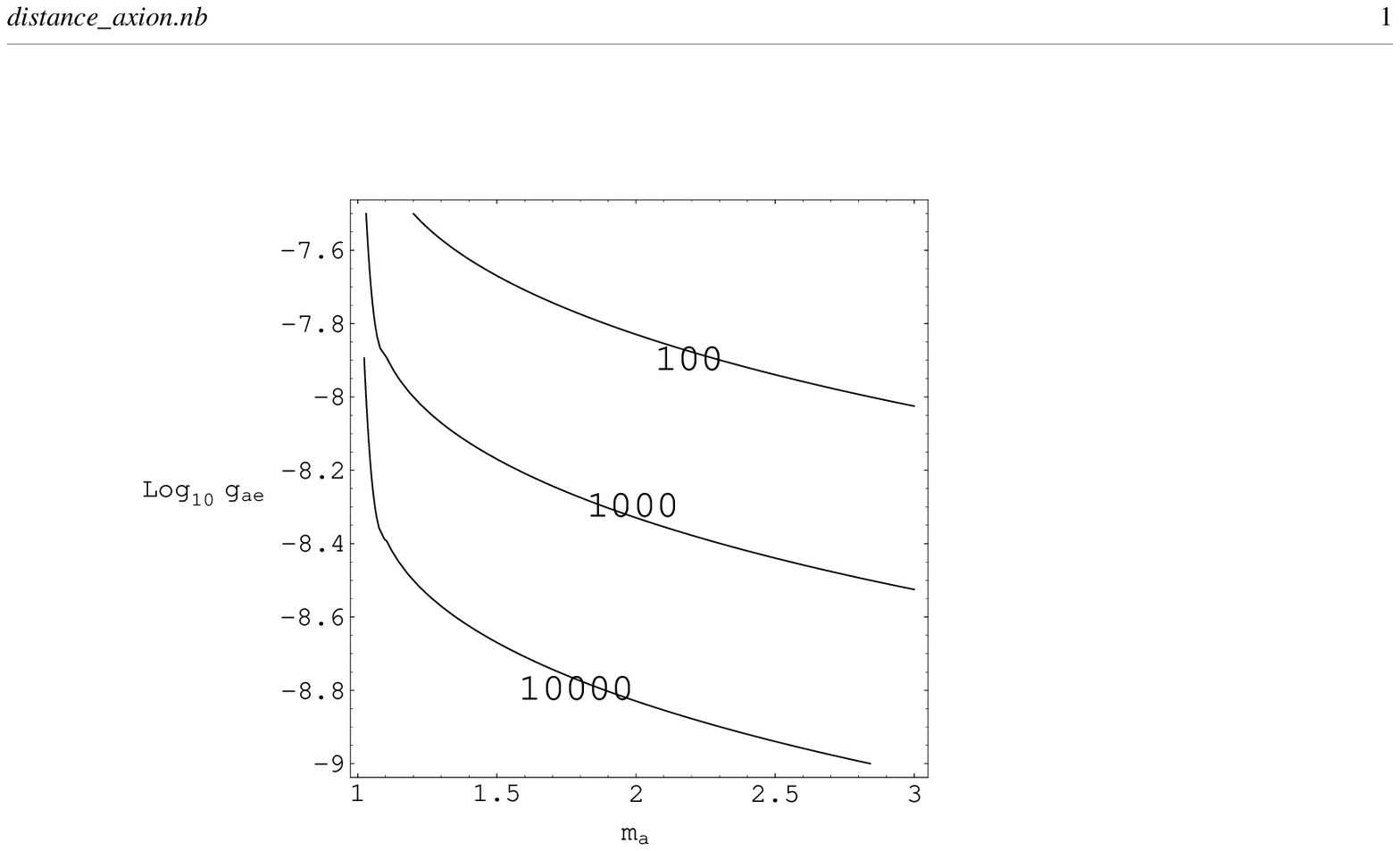}}
}

\begin{center}
\parbox{14cm}{
\caption{$g_{a\gamma}$ and $g_{ae}$ as functions 
of $m_a$[MeV], for the given decay length  
($D_a = 10^2$, 10$^3$ and 10$^4$ km). 
The PGB energy taken is $E=15$ MeV. 
}
}
\end{center}
\end{figure}

\subsubsection*{Mean free path}

The mean free path of a PGB having energy $E$ 
depends on the density $\rho$ and on the temperature $T$ 
of matter, on the neutron and proton mass fractions
$X_n$ and $X_p$ and, of course, on the PGB coupling  
to nucleons $g_{aN}$. (Here and in the following, 
we take for simplicity the PGB couplings to 
proton and neutron to be equal, $g_{ap}=g_{an}=g_{aN}$). 
In ref. \cite{burrows} the mean free path is 
approximated by the following expression:
\be
\lambda_a = g_{aN}^{-2}~F(X_n,X_p,E/T)~l_a(\rho,T), 
\ee
where the coefficient 
$
F = (1+8X_n X_p)^{-1} \left({E/T}\right)
\left(1+{E/ T}\right)^{-1/2} 
$
is of order 1 for typical energies $E\sim 3T$ and
\be
l_a=(2.8\times 10^{-7}~{\rm cm}) \times 
\rho_{12}^{-2}\, T[{\rm MeV}]^{1/2} , 
\label{La}
\ee
with $\rho_{12}$ being the density in units 
of $10^{12}~ {\rm g/cm^3}$.

For a given density and temperature 
one can estimate the critical value of $g_{aN}$ 
for which the PGBs become trapped.  
We denote this value as $g_{aN}^{tr}(\rho,T)$.
Clearly, only an exact knowledge of the density 
and temperature profiles of the system 
can allow a precise determination of this quantity. 
However, we are only interested in a qualitative analysis 
and in our estimate we simply assume that 
the PGB is trapped when $\lambda_a$ 
is smaller than the typical radius of the system $R\sim 50$ km. 
In other words, we define the critical value $g_{aN}^{tr}$ as 
a function of a given density and temperature from 
the following equation\footnote{ Taking into 
account that regions with high densities are smaller than 
$R$, the values we find for $g_{aN}^{tr}$ are clearly 
underestimated with respect to the actual values by a factor  
$\sqrt{R/R(\rho,T)}\sim $ few, where $R(\rho,T)$ is the actual  
size of the region having a given density and temperature. 
However, this approximation is sufficient to our purposes. } 
\be
(g_{aN}^{tr})^{-2}\, l_a(\rho,T) = R.
\label{trap}
\ee
Hence, using eq. (\ref{La}) we obtain: 
\be
g_{aN}^{tr}(\rho,T) = 2.4\times 10^{-7}
\rho_{12}^{-1}
T[{\rm MeV}]^{1/4} , 
\label{g-tr}
\ee 
Assuming for instance $T=5$MeV and $\rho\sim 10^{11}$ g/cm$^3$, 
the PGBs are un-trapped for 
$g_{aN}\le g_{aN}^{tr}=3.6\times 10^{-6}$.
If the density is smaller, $g_{aN}^{tr}$ increases.
E.g., for $\rho\sim 10^{10}$ g/cm$^3$, 
$g_{aN}^{tr}=3.6\times 10^{-5}$, while for $\rho\sim 10^9$ g/cm$^3$,
$g_{aN}^{tr}= 3.6\times 10^{-4}$.

\subsubsection*{Emission rate}

In the conditions of density and temperature typical for  
the central region of the disk, baryonic matter is non-degenerate,
and the PGB production is dominated by brem\-sstrah\-lung from nucleons. 
The analogous process from electrons is less efficient, also because
the latter are partially degenerate at the densities and temperatures 
we are considering here. 
We use for the PGB production rate the expression \cite{raffelt}:
\be
Q_{ND} = 
g_{aN}^{2} R_{ND}(\rho,T) = 
g_{aN}^{2} {272\alpha_\pi^2\over 105 \pi^{3/2}} 
{T^{7/2}\rho^2\over m_N^{9/2}} = 
g_{aN}^2 \rho_{12}^2 T[{\rm MeV}]^{7/2}\times 
2.3\cdot 10^{45} ~{\rm erg/cm}^3{\rm s}
\ee
where, for simplicity, we consider symmetric nuclear matter, i.e. 
$K_{Fn}=K_{Fp}\equiv K_F$, and
$\alpha_\pi\sim 15$ is the pion-nucleon coupling
constant \cite{raffelt}.
In Fig. 2 we show 
${\rm Log}_{10}(R_{ND}$[erg cm$^{-3}$ s$^{-1}])$ as
a function of the temperature and of the density.

\begin{figure}
\begin{center}
\scalebox{0.8}{
\includegraphics*[70,480][400,700]{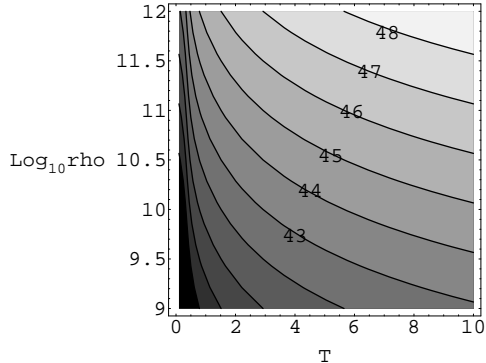}}

\parbox{14cm}{
\caption{ Isocontours for 
${\rm Log}_{10}(R_{ND}$[erg/cm$^{3}$s])  as
a function of the temperature $T$[MeV] and of the density 
$\rho$[g/cm$^3$].
}
}
\end{center}
\end{figure}

Let us now estimate the maximal possible 
luminosity of the PGB emission as a function of $g_{aN}$.    
Clearly, if the PGBs are emitted from a volume  
with given density and temperature in un-trapping regime, 
then their luminosity increases 
with $g_{aN}$ and reaches the maximum at 
$g_{aN}=g_{aN}^{tr}(\rho,T)$.   
For the values of $g_{aN}$ larger than $g_{aN}^{tr}$, 
the PGBs become trapped and their luminosity starts to decrease,  
since only the PGB emission from the surface 
becomes possible. 
To compute the total luminosity we need also to estimate  
a corresponding volume associated with the un-trapped regime. 

The maximal luminosity for a volume $V(\rho,T)$ with 
a given density and temperature reads therefore:
\be
L_a^{max}=
[g_{aN}^{tr}(\rho,T)]^2\times R_{ND}(\rho,T)\times V(\rho,T)
\ee 
We use the results of ref. \cite{ruffert} 
to estimate the magnitude of the maximal luminosity. 
Let us consider, for example, the central zone with  
$\rho \simeq 10^{11}$ g cm$^{-3}$ and $T\simeq 3$ MeV.   
The corresponding volume is roughly $V\sim 10^6$ km$^3$. 
Then, according to eq. (\ref{g-tr}) we obtain 
$g_{aN}^{tr}=3.3\times 10^{-6}$, and thus 
we get $L_a^{max}\sim 10^{56}$ erg/s,
clearly a very large value which would be sufficient to power
the most energetic bursts.\footnote{Certainly, such an enormous 
luminosity is unphysical. The actual luminosity 
should not exceed the rate of the mass transfer which,
for NS-NS merger is 
$dM/dt \simeq 1~ M_\odot$ s$^{-1} = 
1.5\cdot 10^{54}$ erg s$^{-1}$. As we noted above, 
in case of the NS-BH merger the later rate is about 5 times 
bigger.}   
It can be interesting to notice that 
$R_{ND}$ is proportional to $\rho^2$, 
while $g_{aN}^{tr}$ is roughly proportional to $\rho^{-1}$ 
as is indicated by eqs. (\ref{trap}) and (\ref{La}).  
Hence, the maximal luminosity very weakly depends 
on the central density. 

Of course, the actual luminosity can be smaller.   
E.g., for $g_{aN}<g_{aN}^{tr}$
the emission rate is simply proportional to $g_{aN}^2$. 
If, on the other hand, $g_{aN}>g_{aN}^{tr}(\rho,T)$, then  
the emission rate is suppressed due to the PGB trapping. 
In this case the volume in which the PGBs are 
trapped becomes ineffective, since now it contributes 
only through the surface emission, and so the 
dominant contribution comes from outer zones with 
smaller densities in which PGBs are still un-trapped. 
It is difficult to estimate the emission rate in the 
trapping regime. 
Taking, for example, $g_{aN}=3\times 10^{-5}$, i.e. 
one order of magnitude larger then previous value, 
one can roughly estimate the total luminosity as 
$L_a \sim 10^{55}$ erg/s, one order of magnitude 
less then above. 

Therefore, a luminosity larger than ${1\over 10}L_a^{max}\sim
5\times 10^{54}$erg/s can be obtained for $g_{aN}$ in the
range:
\be
10^{-6} < g_{aN} < 3\times 10^{-5} .  
\ee
It is important to notice that, since the PGB luminosity is so
large, a precise determination of the latter is inessential.
Actually, for this range of $g_{aN}$ the PGB emission 
drains all the energy available from the system. 

For sake of simplicity in our analysis we have mainly 
considered the density and temperature profiles which are 
characteristic for the NS-NS merger.
Let us shortly comment also the case of NS-BH merger. 
The latter gives somewhat larger densities and temperatures 
than the NS-NS merger. As observed above, a change in the
density is not affecting the maximal luminosity $L_a^{max}$. 
A slightly larger temperature, on the other hand, 
corresponds to a larger production rate $Q_{ND}$ and therefore 
to a larger luminosity.

\section{The PGB models }

Let us discuss now which particle physics candidate 
could be the PGB with the above properties. As we 
told already, the most familiar example is the axion, 
the PGB associated with the spontaneous breaking of 
the Peccei-Quinn symmetry $U(1)_{PQ}$ needed for the 
solution of the strong CP-problem. The standard axion 
\cite{PQ} is excluded by laboratory experiments long ago. 
However, several types of invisible axions 
have been considered in the literature: 
the DFSZ axion \cite{DFSZ}, the KSVZ or hadronic axion 
\cite{KSVZ}, archion \cite{archion}, etc. In these 
models the PQ symmetry is spontaneously broken due to  
the large VEV of some scalar $S$, singlet of the standard 
model: $\langle S\rangle = f/\sqrt{2}$.  
The axion interactions to matter fields have the form 
(\ref{lagrangian}) where the coupling constants to 
the fermions and photon are related to the
scale $f$ as follows:  
\be
g_{ai}= c_i{m_i\over f} ~~~ (i=e,p,n...), ~~~~~ 
g_{a\gamma} = 
c_\gamma {\alpha \over 2 \pi }{m_N \over f},
\label{couplings}
\ee 
with  $c_i$ and $c_\gamma$ being the 
model dependent coefficients 
(generically, axion models employ a number of 
heavy fermion states that contribute to the colour 
$N$ and electromagnetic $E$ anomalies of the $U(1)_{PQ}$ 
current). 
For any type of axion, its couplings to nucleons do not 
vary strongly from model to model and,  
typically, $c_{p,n}$ are in the range 0.3--1.5.  
As for $g_{ae}$ and $g_{a\gamma}$, their model dependence
is stronger. For example, for the DFSZ axion model
containing only the standard fermion families 
($N=N_f$ is the number of families, and $E/N=8/3$)   
we have 
\be
c_e^{DFSZ}=\sin^2\beta,  ~~~~~
c_{\gamma}^{DFSZ} = {2N_fz\over 1+z},    
\ee 
where $\tan\beta=v_u/v_d$ is the VEV ratio of the up and down 
Higgs doublets $H_u$ and $H_d$.  
Thus, for  $N_f=3$ and a natural range  $v_u/v_d\geq 1$, 
$c_e$ varies from 1/2 to 1, while $c_\gamma$ 
depends on the value of up/down quark mass ratio 
$z=m_u/m_d=0.3-0.7$.  
Therefore, $g_{ae}/g_{a\gamma}=0.47c_e/c_\gamma$ can vary 
between 0.10 and 0.35.  
On the other hand, for the DFSZ-like axion we have 
$g_{aN} = (c_Nm_N/ c_e m_e) g_{ae} \sim (10^3-10^4) g_{ae}$.

For the hadronic axion or the archion the coupling to electrons 
is strongly suppressed. 
For example, for the hadronic axion $g_{ea}=0$ at tree level 
and it emerges from radiative corrections. Namely, we have
\cite{kim}
\be 
c_e = \frac{3\alpha^2}{4\pi} 
\left(E\ln{f\over \Lambda} - c_\gamma 
\ln {\Lambda\over m_e}\right) ,  
~~~~~c_\gamma = E  - {2(4+z)\over 3(1+z)} N ,  
\ee 
where $\Lambda\simeq 1$ GeV is a QCD scale. 
In the GUT context one typically has $E/N=8/3$, 
however this value is not mandatory for a general case. 
Essentially the same order estimates hold for 
the archion couplings, which in addition can have 
also a small tree level coupling to electron, with 
$c_{e}=O(m_e/m_\tau)\sim 10^{-3}$ \cite{archion}.

On the other hand, the axion mass is also related to 
the PQ symmetry breaking scale 
as $m_a \simeq Nm_\pi f_\pi/f$ and thus 
for $f\sim 10^5$ GeV it is too light to be applicable 
for our mechanism. However, 
we do not constrain the magnitude of $m_a$  
with this relation and leave it as free parameter, 
in the range indicated in (\ref{m-limit}). 
In the context of a theory having the $U(1)_{PQ}$ symmetry, 
this would mean that the axion mass is not determined by 
the colour anomaly but rather by some other dynamics. 
For example, one can imagine that its mass  
emerges from the $U(1)_{PQ}$ current anomaly related to 
some hidden gauge sector with the confinement scale larger 
than $\Lambda$ (in this case this axion would solve the strong CP problem 
in this sector).  Alternatively, the Lagrangian could  contain   
the higher order terms cutoff by the Planck scale  
which explicitly break the PQ symmetry 
and thus produce the PGB mass 
\cite{kamionkowski}. In either case, such  
PGB would not be anymore relevant for the strong CP problem. 
One can rather consider it as a "would be" axion which 
could work for strong CP but is prevented to do so by the 
explicit $U(1)_{PQ}$ breaking terms which in general 
have no reason to respect the strong $\theta$ phase orientation.    

For example, taking the Planck scale induced terms in the Lagrangian 
as $\lambda (S^+S)^2S/M_P$, where $\lambda$ is a coupling 
constant, one can estimate 
the PGB mass as \cite{kamionkowski} 
\be
m_a = \sqrt{\lambda}~ 
\left({f\over 10^5~{\rm GeV}}\right)^{3/2} \times 6~{\rm MeV} 
\label{mass}
\ee  
Although the coincidence most probably
is simply accidental, we remark that for the 
DFSZ-like "failed" model with $f={\rm few}\times 10^5$ GeV 
(and hence  
$g_{ae}\sim {\rm few}\times 10^{-9}$,  
$g_{aN} \sim {\rm few}\times 10^{-6}$),  
the PGB mass in eq. (\ref{mass}) 
naturally falls in the range of few MeVs.

\begin{figure}

\parbox{6cm}{
\scalebox{0.6}{
\includegraphics*[70,400][420,720]{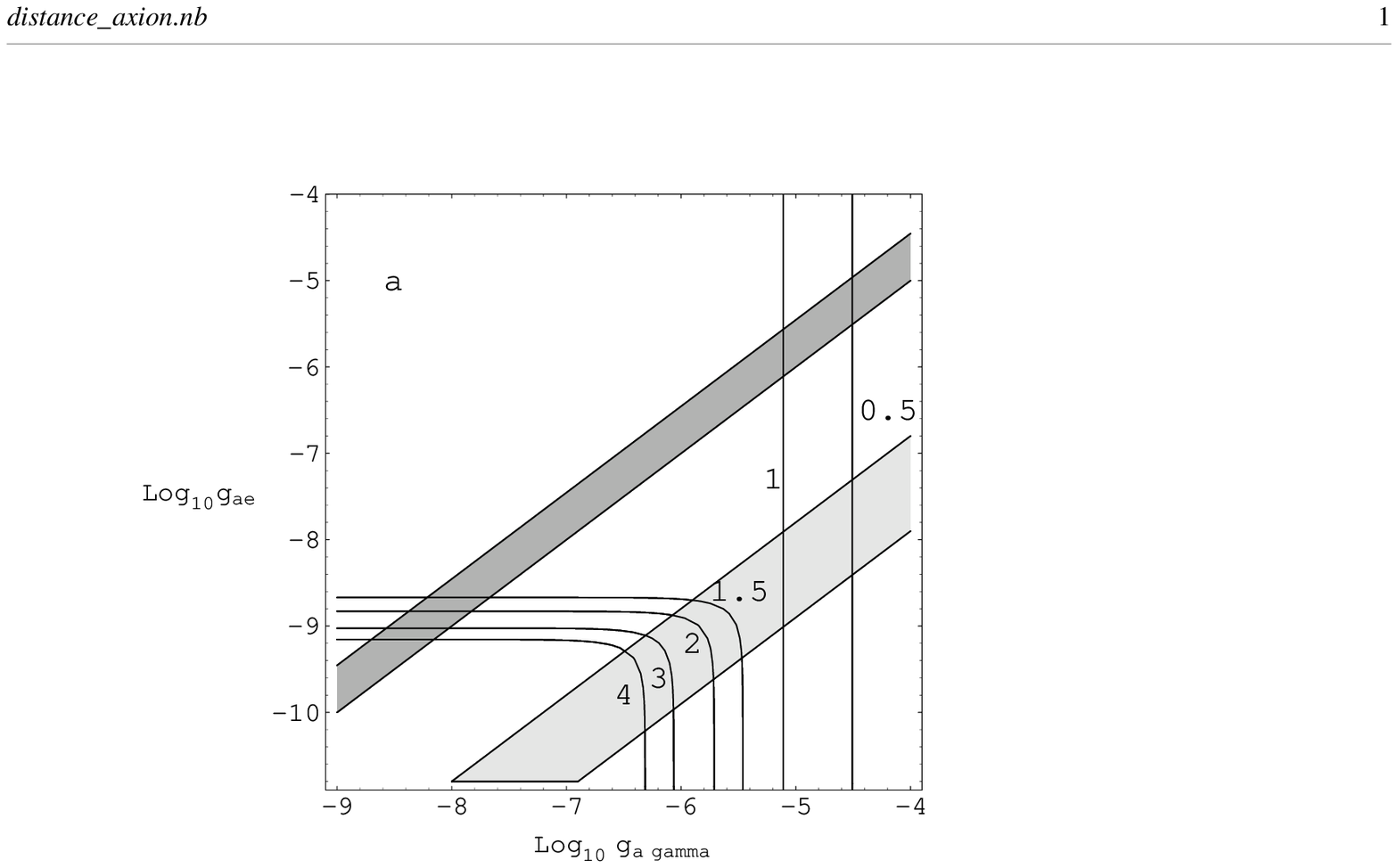}}
}
\nolinebreak
\parbox{6cm}{
\scalebox{0.6}{
\includegraphics*[30,400][420,720]{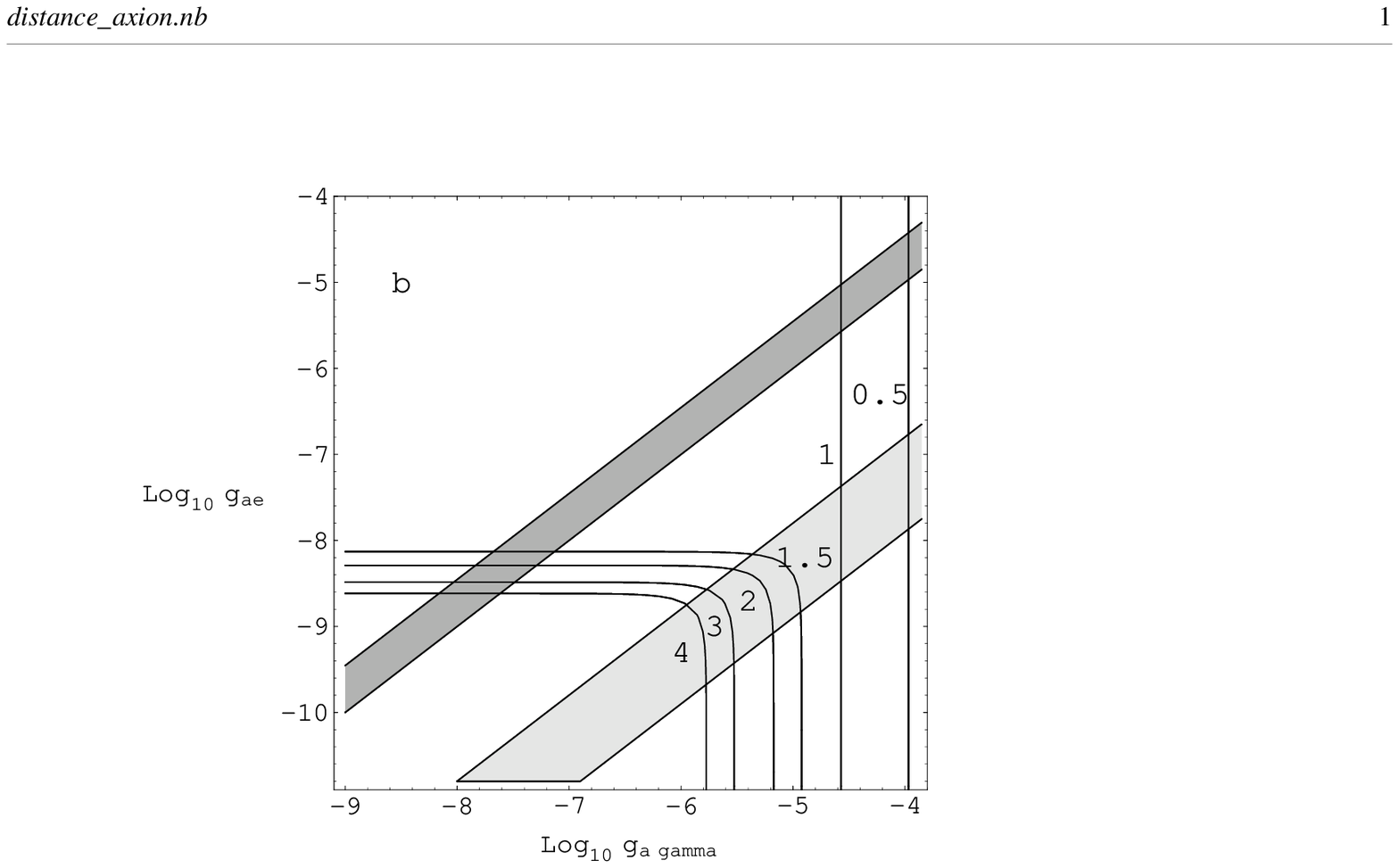}}
}

\begin{center}
\parbox{14cm}{
\caption{Possible values of $g_{ae}$ and $g_{a\gamma}$ 
for a given decay length and axion energy: 
$D_a=10^4$ km, $E$=15 MeV (a) and 
$D_a=500$ km, $E$=9 MeV (b),  for different values
of the mass $m_a$=0.5,1, ..., 4 MeV. 
Shaded areas indicate typical correlations between 
$g_{ae}$ and $g_{a\gamma}$ for 
the DFSZ axion (dark) and hadronic axion or archion (light) 
models.}
}
\end{center}
\end{figure}

We come now back to the analysis of our constraints on the
PGB couplings, interpreting these constraints in the light
of the above discussed models. 
In Fig. 3 we show the isocurves corresponding to a fixed  
decay length $D_a$ of the PGB with 
a given energy $E$,
for different values of $m_a$. We also indicate 
typical correlations between the couplings $g_{ae}$ 
and $g_{a\gamma}$ as they emerge for various axion 
models.  

We see that there can be two type of solutions. 
For the DFSZ axion it is definitely the case 
of decay into $e^+e^-$ which can work. It corresponds 
to values of the constant $g_{ae} \sim {\rm few} 10^{-9}$, 
i.e. $f\sim {\rm few} \times 10^5$ GeV and axion mass of few MeV. 
Interestingly, the parameters range excluded by the 
present reactor experiments \cite{altmann} 
is approaching the parameters spot we are indicating, 
but the parameters value we are suggesting is not yet excluded. 
It must also be remarked that for such values 
of $f$ the axion-nucleon coupling constant lies  
within the range of interest for the GRB, 
$g_{aN} \sim {\rm few} \times 10^{-6}$.

Another solution for the DFSZ, via axion decay into photons 
(vertical lines in Fig. 3 for $m_a < 1$ MeV) 
leads to $g_{ae}>10^{-7}$ which is ruled out by experiment 
\cite{altmann}. 

As for the case of the hadronic axion or archion, 
which have a coupling to electrons suppressed by 3-4 orders 
of magnitude if compared to the DFSZ axion, the decay rate 
into photons is comparable to that into $e^+e^-$ when the latter
is allowed ($m_a > 2 m_e$). 
However, it requires a too large $g_{a\gamma}$, 
above $10^{-7}$. The experimental 
limit of ref. \cite{faissner},
$f>1$ TeV, excludes this possibility. Perhaps, 
this solution is still allowed with a larger axion mass, 
$m_a > 5$ MeV, but this question needs an additional 
investigation.

\section{GRB and Supernovae} 

In the previous sections we have considered GRBs emerging 
by axion emission during a NS-NS or NS-BH merger. However, 
the proposed mechanism can have strong implications also 
for the GRBs associated with supernova explosion, as well 
as for the supernova explosion itself. 

In the typical situation we considered, axions produced 
in a compact disk fly away 
taking the big portion of the available energy. 
In case of merger the typical size of the system 
is of the order of 100 km -- there is essentially no baryonic 
matter outside this radius, so the axion flux, after 
decay into photons or $e^+e^-$ at 
a distance larger then 100 km, converts 
into a fireball -- a bubble (or jet) 
of hot relativistic plasma expanding with a 
large Lorentz factor.\footnote{
A possible speculation concerns the observed spectrum of GRBs. 
The latter display a characteristic paucity for energies below 
few tens of keV. Moreover it typically has a maximum 
(the hardness H \cite{band}) 
for energies of the order of few hundreds keV. 
Although it is not clear if there is any relation between 
the ``internal engine'' and the observed spectrum, 
we cannot avoid to remark a similarity between the typical 
value of the hardness and the mass of the PGB we have 
discussed in this paper.} 

However, the axions can be produced also in the core 
of collapsing stars. In particular, for 
$g_{aN} \sim 10^{-6}$ the axions are essentially in 
the trapping regime in the supernovae exploding via core collapse.
Axions are  
emitted from the axiosphere, having a temperature 
decreasing with time from $T\simeq 4$ MeV at $t=100$ ms to  
$T\simeq 3$ MeV at $t=1$ s. The corresponding 
luminosities are 
$L_a \sim 3\cdot 10^{51}$ erg/s at $t\sim 100$ ms and 
$L_a \sim 0.5\cdot 10^{51}$ erg/s at $t\sim 1$ s 
\cite{burrows}. Therefore, in total an energy $\sim 10^{51}$ erg can be 
emitted during the collapse period and subsequent 
cooling of the proto-neutron star, in terms of axions 
having an average energy $E\sim 10$ MeV. 
Since $E$ is of the same order as the one 
in case of mergers, the value 
$D_a=c\tau\gamma$ is again of the order of few 
hundred or few thousand km.  
Then in our scenario the impact of the axions 
simply depends on the geometrical size of the collapsing star. 

In particular, supernovae type Ib,c are the result of 
core collapse of relatively small stars, where the 
hydrogen and perhaps also the helium shells are missing.
A typical radius for SN Ic is
$R\sim 10^4$ km. Therefore, the axion decay length 
$D_a=c~\tau\gamma$ is comparable to $R$. 
This in turn implies that 
$\exp (-R/D_a)$ is not very small, it can be of order 0.1 to 1, 
and thus a reasonable amount of axions decay outside the mantle. 
Clearly, this also implies that the bulk of axions 
decaying outside the star produces a fireball 
(with a Lorentz factor $\Gamma~ \sim 3~T/2~m_e$).
We can conclude that, associated with a supernova type Ib,c
a GRB could take place, having a typical energy 
$\sim 10^{51}$ erg. 

In case of SN type II, which are associated with large stars,  
e.g. red giants, having an extended hydrogen shell 
($R\sim (1-20)\times 10^7$ km),  
the axion decay takes place completely inside the mantle 
and thus cannot be observed as GRB. For example, for 
SN 1987A, related to the blue giant Sanduleak having 
$R=3\cdot 10^7$ km, the fraction of axions decaying 
outside the envelope, $\exp(-R/D_a)$, is essentially zero. 
Nevertheless, it can be of interest to think that the 
energy of order $10^{51}$ erg released after axion decay 
at distances of few hundred or thousand km can help 
to solve the painful problem of mantle ejection and therefore of
supernova explosion (in the prompt mechanism, shock stalls 
at a distance of few hundred km).\footnote{
The possibility that the two photon decay of the axion-like 
particles with mass $0.15-1$ MeV could provide a mechanism 
whereby gravitationally collapsing massive stars may eject 
their outer mantles and envelopes in supernova explosions 
was first considered in ref. \cite{schramm}. It should be 
noted, however, that the range for the coupling 
$g_{a\gamma}$ explored in ref. \cite{schramm}  
actually corresponds to the vertical band of our Fig. 3 
and it is strongly excluded by the experimental limits 
\cite{faissner}. 
We thank S. Blinnikov for driving our attention 
to the ref. \cite{schramm}.}

\section{Summary and conclusions}

We have shown that it exists a range of values for the PGB 
parameters such that a strong PGB emission takes place during 
the NS-NS or NS-BH merger.
Choosing appropriately the parameters' value, the total luminosity 
for the PGB emission
can be as large as ${\rm few}\times 10^{55}$ erg/s. 
Moreover, at variance with
the mechanism based on $\nu\bar\nu\to e^+e^-$ annihilation 
which has a small (about one percent) efficiency, 
the PGB emission mechanism suffers no energy deficit:  
all emitted PGBs, due to their decay into 
$e^+e^-$ or photons,  
convert into the ultrarelativistic plasma, the fireball,  
which is at the origin of the burst. 
This plasma can be produced at distances 
as large as 100 km $< D_a <$ 10000 km 
(which are difficult to achieve in the case of plasma produced via 
$\nu\bar\nu$ annihilation), is not contaminated by baryonic matter 
and thus a large Lorentz factor can be obtained. 
Our conclusion is supported by the numerical simulations 
given in the recent ref. \cite{aloy}.

An interesting possibility opened by our mechanism is the 
association of ``weak'' GRBs with the supernovae type Ib,c. 
In addition, the PGB emission could also help to solve the problem of
supernovae type II explosion, thus providing an unified 
theoretical base for the GRB and SN phenomena. 

The needed parameter range for such a PGB indicate 
an axion-like particle having a mass $m_a$ of a few MeV, 
coupling to nucleons $g_{aN}\sim {\rm few}\times 10^{-6}$  
and to electrons $g_{ae}\sim {\rm few}\times 10^{-9}$. 
This range of coupling constants coincides to that of the 
invisible axion with the Peccei-Quinn symmetry 
breaking scale $f\sim {\rm few}\times 10^5$ GeV. However, 
for such scale $f$ a "true" axion would have a mass 
of few eV whereas we our PGB has a mass of few MeV, 
i.e. about a million times larger. 
Present experimental data and astrophysical limits  
cannot exclude this particle, however the relevant parameters' 
window is not far from the present experimental 
limits and it can be experimentally tested in the close
future.

\section*{Acknowledgements}

We would like to thank Sergei Blinnikov and Filippo Frontera 
for reading the manuscript and illuminating conversations.  
Useful discussions with Venya Berezinsky, 
Enrico Montanari, Amedeo Tornamb\'e 
and Mario Vietri are also gratefully acknowledged.

\end{document}